# MULTIWAVELENGTH OBSERVATIONS OF A BRIGHT IMPACT FLASH DURING THE JANUARY 2019 TOTAL LUNAR ECLIPSE

José M. Madiedo[1], José L. Ortiz[2], Nicolás Morales[2], Pablo Santos-Sanz[2]

[1] Facultad de Ciencias Experimentales, Universidad de Huelva. 21071 Huelva (Spain).
[2] Instituto de Astrofísica de Andalucía, CSIC, Apt. 3004, Camino Bajo de Huetor 50, 18080 Granada, Spain.

ABSTRACT

We discuss here a lunar impact flash recorded during the total lunar eclipse that occurred on 2019 January 21, at 4h 41m 38.09 ± 0.01 s UT. This is the first time ever that an impact flash is unambiguously recorded during a lunar eclipse and discussed in the scientific literature, and the first time that lunar impact flash observations in more than two wavelengths are reported. The impact event was observed by different instruments in the framework of the MIDAS survey. It was also spotted by casual observers that were taking images of the eclipse. The flash lasted 0.28 seconds and its peak luminosity in visible band was equivalent to the brightness of a mag. 4.2 star. The projectile hit the Moon at the coordinates 29.2 ± 0.3 ºS, 67.5 ± 0.4 ºW. In this work we have investigated the most likely source of the projectile, and the diameter of the new crater generated by the collision has been calculated. In addition, the temperature of the lunar impact flash is derived from the multiwavelength observations. These indicate that the blackbody temperature of this flash was of about 5700 K.

1. INTRODUCTION

The Earth and the Moon continuously experience the impact of meteoroids that intercept the path of both celestial bodies. The analysis of these collisions provides very valuable data that allows us to better understand the Earth-Moon meteoroid environment. The study of meteoroid impacts on the Moon from the analysis of the brief flashes of light that are generated when these particles hit the lunar ground at high speeds has proven to be very useful to investigate this environment. For instance, the analysis of the frequency of these events can provide information about the impact flux on Earth (see e.g. Ortiz et al. 2006; Suggs et al. 2014; Madiedo et al. 2014a, 2014b). Also the initial kinetic energy of the projectile, its mass, and the size of the resulting crater can be obtained. For events produced by large (cm-sized or larger) particles, one of the main benefits of this technique over the systems that analyze meteors produced by the interaction of meteoroids with the atmosphere of our planet is that a single instrument covers a much larger area on the lunar surface (typically of an order of magnitude of $10^6$ km$^2$) than that monitored in the atmosphere of the Earth by a meteor-observing station.

The monitoring of lunar impact flashes by means of telescopes and high-sensibility cameras dates back to the 1990s. Since the first systematic observations performed by Ortiz et al. (1999) in this field, different authors have obtained information about the collision with the lunar surface of meteoroids from several sources. Thus, flashes associated with impactors belonging to the sporadic meteoroid background and to different meteoroid streams have been recorded and described (see for instance Madiedo et al. 2019 for a comprehensive review about this topic). Some synergies have been found when this method is employed in conjunction with the technique based on the monitoring and analysis of meteors produced by meteoroids entering the atmosphere (Madiedo et al. 2015a,b). Even fresh impact craters associated to observed lunar impact flashes have been also observed by means of the Lunar Reconnaissance Orbiter (LRO) probe, which is in orbit around the Moon since 2009 (Robinson et al. 2015, Madiedo and Ortiz 2018, Madiedo et al. 2019). More recently, since 2015, lunar impact flashes observations simultaneously performed in several spectral bands allowed us to estimate the temperature of impact plumes (Madiedo and Ortiz 2016; Madiedo et al. 2018; Bonanos et al. 2018).

Despite its multiple advantages, this technique has also some important drawbacks, since the results are strongly dependent on the value given to the luminous efficiency. This parameter is the fraction of the kinetic energy of the projectile emitted as visible light as a consequence of the collision. The value of the luminous efficiency is not known with enough accuracy. The comparison between the calculated size of fresh craters associated to observed impact flashes and the experimental size measured by probes orbiting the Moon can play a fundamental role to better constrain the value of this efficiency (Ortiz et al. 2015).

Another drawback of this technique is related to the fact that, since most of these flashes are very dim, they must be recorded against a dark background. For this reason, the method is based on the monitoring of the nocturnal region of the Moon. The area directly illuminated by the Sun must be avoided in order to prevent the negative effects of the excess of scattered light entering the telescopes. This implies that, weather permitting, the monitoring by means of telescopes of these flashes is limited to those periods where the illuminated fraction of the lunar disk ranges between about 5% and 50-60%, i.e., about 10 days per month during the waxing and waning phases (Ortiz et al. 2006, Madiedo et al. 2019). Lunar eclipses provide another opportunity to monitor lunar impact flashes out of this standard observing period, since during these the Moon gets dark. However, because of the typical duration of lunar eclipses, this extra observational window is relatively short when compared to a standard observing session. Besides, the possibility to detect dimmer impact flashes, which are more frequent than brighter ones, depend on the intrinsic brightness of the eclipse, which in turn depend on the aerosol content at stratospheric levels. In general, the lunar ground is brighter in visible light during a lunar eclipse than the lunar ground in standard observing periods during the waning and waxing phases. These factors, which pose some difficulties to the detection of lunar impact flashes, might have contributed to the fact that, despite several researchers have conducted impact flashes monitoring campaigns during lunar eclipses, no team succeeded until now. The first lunar impact flash monitoring campaign performed by our team during a total lunar eclipse was conducted by the second author of this work in October 2004. In 2009, the pioneer survey developed by Ortiz et al. (1999) was renewed and named Moon Impacts Detection and Analysis System (MIDAS) (Madiedo et al. 2010; Madiedo et al. 2015a, 2015b). This project is conducted from three astronomical observatories located in the south of Spain: Sevilla, La Sagra and La Hita (Madiedo and Ortiz 2018, Madiedo et al. 2019). In this context, our survey observed a flash on the Moon during the total lunar eclipse that took place on 2019 January 21. This flash was also spotted by casual observers that were taking images of this eclipse, or streaming it live on the Internet (https://www.reddit.com/r/space/comments/ai79zy/possible_meteor_impact_on_moon_during_the_eclipse/). The MIDAS survey was the first to confirm that this flash was generated as a consequence of the collision of a meteoroid with the lunar soil at high speed, so that this is the first lunar impact flash ever recorded during a lunar eclipse and discussed in the scientific literature. The news was covered by communication media all around the world. From a scientific point of view, it offered the opportunity to monitor the Moon with an angular orientation very different to that of the regular campaigns at waxing and waning phases and it was a good opportunity to test new equipment for the monitoring of lunar impact flashes, and provided valuable data in relation to the study of impact processes on the Moon. We focus here on the analysis of this impact event.

2. OBSERVATIONAL TECHNIQUE

The impact flash discussed in this work was observed from Sevilla on 2019 January 21. Our systems at the observatories of La Sagra and La Hita could not operate because of adverse weather conditions. In Sevilla, five f/10 Schmidt-Cassegrain telescopes were used. Two of these instruments had an aperture of 0.36 m, and the other three telescopes had a diameter of 0.28 cm. These telescopes employed a Watec 902H Ultimate video camera connected to a GPS-based time inserter to stamp time information on each vide frame. The configuration of these cameras, which are sensitive in the wavelength range between, approximately, 400 and 900 nm, is explained in full detail in Madiedo et al. (2018). The observational setup consisted also of two 0.10 m f/10 refractors endowed with Sony A7S digital cameras, which provided colour imagery and employ the IMX235 CMOS sensor. One of these was configured to take still images each 10 s with a resolution of 4240x2832 pixels, while the other recorded a continuous video sequence of the eclipse at 50 fps with a resolution of 1920x1080 pixels. A third Sony A7S camera working in video mode was attached to a Schmidt-Cassegrain telescope with an aperture of 0.24 m working at f/3.3. However, because of a technical issue that occurred during the eclipse, this telescope could not be finally operated. The Sony A7S cameras are sensitive within the wavelength range between, approximately, 400 and 700 nm. These have been used in the framework of our survey for the first time during this monitoring campaign to take advantage of the colour information they could provide. Also, the larger field of view of these instruments allowed for a full coverage of the lunar surface during the totality phase of the eclipse, in contrast with the Schmidt-Cassegrain telescopes with the Watec cameras, which can monitor only an area of the Moon of around $4 \cdot 10^6$ to $8 \cdot 10^6$ km$^2$ (see for instance Madiedo et al. 2015a,b and Ortiz et al. 2015).

No photometric filter was attached to the cameras employed with the 0.36 m and two of the 0.28 m Schmidt-Cassegrain telescopes. These provided images in the wavelength range between, approximately, 400 and 900 nm. The third 0.28 m SC telescope employed a Johnson-Cousin I filter. Observations performed with the two refractors were also unfiltered.

We did not focus on the monitoring of any particular region on the lunar disk. Instead, our telescopes were aimed so that the whole lunar disk was monitored during the totality phase of the eclipse, with each instrument covering a specific area of the lunar surface, and with at least two instruments monitoring a common area. Before and after the totality, the region of the Moon not occulted by the Earth's shadow was avoided. The MIDAS software (Madiedo et al. 2010, 2015a) was employed to automatically detect lunar impact flashes in the images obtained with the above-mentioned instrumentation.

3. OBSERVATIONS

Our lunar monitoring campaign took place on 2019 January 21 from 3h 33m UT to 6h 50m UT. These times correspond to the first and last contact with the Earth's umbra, respectively. Excellent weather conditions allowed us to monitor the Moon during the whole time interval, so the effective observing time of was of 3.2 hours. This resulted in the detection of a flash at 4h 41m 38.09 ± 0.01 s UT (Figure 1), about 21 seconds after the totality phase of the eclipse began. This event, which lasted 0.28 s, was simultaneously recorded by two of our instruments: one of the 0.36 m Schmidt-Cassegrain telescopes, and the 0.1 m refractor with the Sony A7S camera that recorded the continuous video sequence of the eclipse. This flash was also reported in social networks by several observers at different locations in Europe, America and Africa (https://www.reddit.com/r/space/comments/ai79zy/possible_meteor_impact_on_moon_during_the_eclipse/). The MIDAS team confirmed that it was associated with an impact event on the Moon. Table 1 contains the main parameters derived for this impact flash. By means of the MIDAS software (Madiedo et al. 2015a, 2015b) we determined that the impactor hit the Moon at the selenographic coordinates 29.2 ± 0.3 ºS, 67.5 ± 0.4 ºW, a position close to crater Lagrange H. This is located next to the west-south-west portion of the lunar limb.

It is worth mentioning that astronomers at the Royal Observatory in Greenwich reported a second flash at 4:43:44 UT (Emily Drabek-Maunder, personal communication). We tried to locate this flash in our recordings by checking them automatically with our MIDAS software. We also checked them manually, by performing a visual inspection of the videos frame by frame. We allowed for a timing uncertainty of around 1 minute, which is well above the 5 seconds time difference between the time reported by this observatory for the first flash (4:41:43 UT) and the time specified by our GPS time inserters. However, this event was not present in any of the images recorded by our systems and, to our knowledge, no other casual observer spotted it. This means that it should have been produced by a different phenomenon, and not by a meteoroid hitting the lunar ground. The MIDAS survey uses at least two instruments monitoring the same lunar area in order to have redundant detection to discard false positive impact flashes due to cosmic ray hits, satellite glints and other possible phenomena that may mimic the impact flashes.

## 4. RESULTS AND DISCUSSION

### 4.1. Impactor source

Since the technique employed to detect lunar impact flashes cannot unambiguously provide the source of the impactors that produce these events (Madiedo et al. 2015a, 2015b, 2019), we have followed the approach described in (Madiedo et al. 2015a, 2015b) to determine the most likely source of the meteoroid that generated the flash discussed here.

The observing date did not coincide with the activity period of any major meteor shower on our planet and so the impactor should be associated either with a minor meteoroid stream or with the sporadic meteoroid component. Our meteor stations, which operate in the framework of the SMART project (Madiedo 2014, 2017), recorded that night meteors from the January Comae Berenicids (JCO), the δ-Cancrids (DCA), and the ρ-Geminids (RGE), but the activity of all of these corresponded to a zenithal hourly rate (ZHR) < 1 meteor/h. Besides, the geometry for the impact of the DCA and RGE streams did not fit that of the lunar impact flash: these meteoroids could not hit the lunar region where the flash was recorded. So, we considered the sporadic background and the JCO meteoroid stream as potential sources of the event. The association probabilities corresponding to these sources, labelled as $p_{SPO}$ and $p_{JCO}$, respectively, were obtained by following the technique developed by Madiedo et al. (2015a, 2015b). Thus we have calculated $p_{JCO}$ with our software MIDAS, which obtains this probability from Equation (15) in the paper by Madiedo et al. (2015b). In this calculation the zenithal hourly rate and the population index of the January Comae Berenicids have been set to 1 meteor/h and 3, respectively, and HR=10 meteors/h was set for the activity of the sporadic component (see for instance Dubietis and Arlt, 2010). From this analysis $p_{JCO}$ yields 0.01, with $p_{SPO} = 1 - p_{JCO} = 0.99$. According to this, the probability that the impactor is linked to the sporadic meteoroid component is of about 99%. In these calculations an average impact velocity and an impact angle of sporadics on the Moon of 17 km s$^{-1}$ and 45º, respectively, have been assumed (Ortiz et al. 1999). For impactors associated with the JCO meteoroid stream this velocity was set to 65 km s$^{-1}$ (see e.g. Jenniskens 2006) and, according to the impact geometry, the angle of impact would be of around 54º in this case.

4.2. Impactor kinetic energy and mass

We recorded the impact flash with the Watec camera in white light only. Since no observations with different photometric filters were available for this CCD device, we could not employ color terms for the photometric analysis of the event. As explained in the next section, color terms could be employed in the case of the Sony A7S camera. So, as in previous works (see, e.g., Ortiz et al. 2000, Yanagisawa et al. 2006, Madiedo et al. 2014), the brightness of the flash as recorded with the Watec camera was estimated by comparing the luminosity of this event with the known V magnitude of reference stars observed with the same instrumentation at equal airmass. In this way we could determine that the peak magnitude of the impact flash was 4.2 ± 0.2. Figure 2 shows the lightcurve of the flash as recorded by means of the 0.36 m telescope that spotted the event. Using t=0.28 in the empiric equation

$$t = 2.10 \exp(-0.46 \pm 0.10 \, m) \tag{1}$$

that links impact flash duration t and magnitude m (Bouley et al. 2012), we come up with a 4.1 mag for the flash, which is close to the derived 4.2 mag.

The calculations in this section are performed from the data collected by this instrument, since its larger aperture and the higher sensitivity of its CCD camera allowed us to record the evolution of the impact flash in much more detail than with the 0.1 m refractor. This refractor telescope just registered the peak luminosity of the flash and so the lightcurve of the event cannot be constructed from its recordings.

As explained in detail in Madiedo et al. (2018), the energy radiated on the Moon by the flash can be obtained from the integration of the power radiated by the event:

$$P = 3.75 \cdot 10^{-8} \cdot 10^{(-m/2.5)} \pi f \Delta \lambda R^2 \qquad (2)$$

Here the magnitude of the flash varies with time according to the lightcurve of the event, and f quantifies the degree of isotropy of the emission of light. Since we have considered that light was isotropically emitted from the lunar ground, we have set f = 2 (Madiedo et al. 2018). The distance between our observatory on Earth and the impact location on the Moon at the instant when the event took place was R= 364831.2 km. For the wavelength range Δλ corresponding to the luminous range we have set Δλ = 0.5 μm (see for instance Ortiz et al. 2000 and Madiedo et al. 2019).. By entering these parameters in Eq. (2) the energy radiated on the Moon yields E = (1.96±0.39)·$10^7$ J.

This radiated energy is a fraction of the kinetic energy $E_k$ of the meteoroid. That fraction is called the luminous efficiency, which is wavelength-dependent and is usually denoted by η (Bellot Rubio et al. 2000a, 2000b; Ortiz et al. 2000; Madiedo et al. 2018, 2019):

$$E = \eta\, E_k \qquad (3)$$

Since the value of the radiated energy derived from Eq. (2) depends on the wavelength range considered, the luminous efficiency for that same spectral range defined Δλ by must be employed. On the contrary, we would arrive to the non-sense conclusion that the kinetic energy of the projectile would be also a function of the spectral range, instead of depending only on the mass and velocity of the projectile. The concept "luminous" refers to the above-mentioned luminous range, and it was defined to correspond to the range of sensitivity of typical CCD detectors (i.e., from around 400 to about 900 nm) used in the first works on lunar impact flashes and luminous efficiencies (see e.g. Bellot-Rubio et al. 2000a, 2000b; Ortiz et al. 2000; Yanagisawa et al. 2006). Other wavelength ranges can be of course defined and employed, but this consistency between Δλ, E and η must be maintained. For other spectral ranges the fraction of the kinetic energy of the impacting meteoroid converted into radiation in the corresponding photometric bands should be denoted by using subscripts, such as $\eta_R$, for the R-band, $\eta_I$ for the I-band, etc., to avoid confusing it with η (Madiedo et al. 2018, 2019). In previous works the value employed for the luminous efficiency was $\eta = 2 \cdot 10^{-3}$ (Ortiz et al. 2006, 2015). However, this value was derived by assuming f=3 for the degree of isotropy factor (see, for instance, Ortiz et al. 2006). Since in this work we have considered f=2, we have to multiply this value of the efficiency by 3/2, as explained in Madiedo et al. (2018). As a consequence of this, the value considered for η in the luminous range for the flash yields η = 3·10$^{-3}$. In this way, the kinetic energy $E_k$ of the impactor is $E_k$ = (6.55±0.63)·10$^9$ J. The impactor mass M derived from this kinetic energy is M = 45 ± 8 kg for a sporadic meteoroid impacting at velocity of 17 km s$^{-1}$. Its size is readily obtained from the bulk density of the particle. The average value of this bulk density for projectiles associated with the sporadic meteoroid background is $ρ_P$=1.8 g cm$^{-3}$ according to Babadzhanov and Kokhirova (2009). This density yields a diameter for the impactor $D_P$ = 36 ± 2 cm. However, if the projectile consisted of soft cometary materials, with a bulk density of 0.3 g cm$^{-3}$, or ordinary chondritic materials, with $ρ_P$ = 3.7 g cm$^{-3}$ (Babadzhanov and Kokhirova 2009), the size of the projectile would yield $D_P$ = 66 ± 4 cm and $D_P$ = 29 ± 2 cm, respectively.

4.3. Temperature of the impact plume

Unfortunately, the impact flash was not recorded by the 0.28 m telescope with the Johnson I filter, since the event took place outside the field of view of this instrument. So, we could not derive the temperature of the impact flash by comparing the energy flux density measured in the luminous and the I ranges (Madiedo et al. 2018). Instead, we followed here a different approach on the basis of the colour images recorded by the 0.1 m refractor and the Sony A7S camera. The decomposition of these colour images into its individual R, G and B channels (Figure 3) provides a multiwavelength observation of the impact flash, which can be employed, for instance to derive the flash temperature, assuming blackbody emission. To do so, we have performed a photometric calibration of the Sony A7S camera to derive the flash magnitude in the Johnson-Cousins R, V and B bands from its measured luminosity in R, G and B channels of the video stream. For this conversion color term corrections are necessary. It is worth mentioning that the Sony A7S camera has a built-in NIR blocking filter, but in the spectral response of the device, no leakage in the NIR was observed. The calibration procedure has been performed as follows.

The magnitudes $m_R$, $m_V$ and $m_B$ in the Johnson-Cousins photometric system are given by the following standard relationships:

$$m_R = r + ZP_R + (m_V - m_R) C_R - K_R A \qquad (4)$$
$$m_V = v + ZP_V + (m_V - m_R) C_V - K_V A \qquad (5)$$
$$m_B = b + ZP_B + (m_B - m_V) C_B - K_B A \qquad (6)$$

In these equations $ZP_R$, $ZP_V$, and $ZP_B$ are the corresponding zero points for each photometric band, $K_R$, $K_V$, and $K_B$ are the extinction coefficients, and A is the airmass; r, v, and b are the instrumental magnitudes in R, V and B band, and are defined by

$$r = -2.5\log(S_R) \quad (7)$$
$$v = -2.5\log(S_G) \quad (8)$$
$$b = -2.5\log(S_B) \quad (9)$$

where $S_R$, $S_G$, and $S_B$ are the measured signals. We employed 30 calibration stars within the Messier 67 open cluster, with known $m_R$, $m_V$ and $m_B$, to obtain the value of the color terms $C_R$, $C_V$, and $C_B$ and the coefficients $ZP_R$, $ZP_V$, $ZP_B$, $K_R A$, $K_V A$, and $K_B A$ by performing a least-squares fit (Figures 4 to 6). These stars were observed with the same refractor telescope and Sony A7S camera employed to record the flash. Their signals $S_R$, $S_G$ and $S_B$ were measured by performing an aperture photometry. Since the calibration stars and the impact flash were observed at the same airmass, the least-squares fit provided the sum of ZP and KA in a single constant for each band R, V and B. The values resulting from this fit are shown in Table 2. By inserting in Eqs (4-6) the measured flash signals in R, G and B channels, the peak magnitude of the flash in R, V and B bands yield, respectively, $m_R = 3.53 \pm 0.19$, $m_V = 4.08 \pm 0.10$ and $m_B = 4.75 \pm 0.09$. The value calculated for $m_V$ fits fairly well the $4.2 \pm 0.2$ magnitude in V band derived from the images obtained with the Watec camera.

From these magnitudes, the energy flux densities observed on our planet for the above-mentioned bands (labelled as $F_R$, $F_V$, and $F_B$) have been estimated by employing the following equations:

$$F_R = 1.80 \cdot 10^{-8} \cdot 10^{(-m_R/2.5)} \quad (10)$$

$$F_V = 3.75 \cdot 10^{-8} \cdot 10^{(-m_V/2.5)} \quad (11)$$

$$F_B = 6.70 \cdot 10^{-8} \cdot 10^{(-m_B/2.5)} \quad (12)$$

where the multiplicative constants $1.80 \cdot 10^{-8}$, $3.75 \cdot 10^{-8}$ and $6.70 \cdot 10^{-8}$ correspond to the irradiances, in $Wm^{-2}\mu m^{-1}$, for a mag. 0 star in the corresponding photometric band. The effective wavelengths for these bands are $\lambda_R = 0.70$ µm, $\lambda_V = 0.55$ µm, and $\lambda_B = 0.43$ µm, respectively. These parameters have been provided by the magnitude to flux converter tool of the Spitzer Science Center (http://ssc.spitzer.caltech.edu/warmmission/propkit/pet/magtojy/). The flux densities given by Eqs (10-12) are plotted in Figure 7. By assuming that the flash behaves as a blackbody, these flux densities have been fitted to Planck's radiation law. The best fit is obtained for T = 5700 ± 300 K. This temperature agrees with the statistics of flash temperatures derived with 2-color measurements from the Neliota survey, for which blackbody temperatures ranging between 1300 and 5800 K have been estimated (Avdellidou and Vaubaillon 2019). Our result is in the high-end tail of the blackbody temperature flash distribution shown in Avdellidou and Vaubaillon (2019) from a sample of 55 impact flashes with magnitudes in R band ranging between 6.67 to 11.80. Lower temperatures can be fit to our data by assuming optically thin emission modulated by the optical depth, but we cannot determine the optical depth of the emitting hot cloud at different wavelengths without making too many assumptions. When observations at 4 or more wavelengths become available we will be able to shed more light on this.

4.4. Crater size and potential observability by lunar spacecraft

The estimation of the size of fresh craters associated with observed lunar impact flashes is fundamental to allow for a better constraint of the luminous efficiency, a key parameter which is not yet known with enough accuracy. Thus, if these craters are later on observed and measured by probes in orbit around the Moon, the comparison between predicted and experimental sizes is of a paramount importance to test the validity of the parameters and theoretical models employed to analyze these impacts. Different models, which are also called crater-scaling equations, can be employed to estimate the size of these fresh craters, and most studies in these field employ either the Gault model or the Holsapple model. The Gault equation is given by the following relationship (Gault, 1974):

$$D = 0.25 \rho_p^{1/6} \rho_t^{-0.5} E_k^{0.29} (\sin\theta)^{1/3} \qquad (13)$$

D is the rim-to-rim diameter, $\rho_p$ and $\rho_t$ are the projectile and target bulk densities, respectively, and the angle of impact $\theta$ is measured with respect to the local horizontal (Melosh, 1989). We have employed $\theta=45º$ for sporadic meteoroids, and for the target bulk density we have considered $\rho_t = 1.6$ g cm$^{-3}$. By entering in this model the previously-obtained value of the kinetic energy $E_k$, the diameter D for impactor bulk densities $\rho_p$ of 0.3, 1.8 and 3.7 g cm$^{-3}$ yields 10.1 ± 0.5 m, 13.6 ± 0.6 m, and 15.3 ± 0.7 m, respectively.

We have also derived the crater size from the following equation, which was proposed by Holsapple (1993):

$$D = 2.6 K_r \left[ \frac{\pi_v M}{\rho_t} \right]^{1/3}. \tag{14}$$

D is again the rim-to-rim diameter, and $\pi_v$ is an adimensional factor which has the following form:

$$\pi_v = K_1 \left[ \left( \frac{ga}{(V\sin(\theta))^2} \right) \left( \frac{\rho_t}{\rho_P} \right)^{\frac{6\nu-2-\mu}{3\mu}} + \left[ K_2 \left( \frac{Y}{\rho_t (V\sin(\theta))^2} \right) \left( \frac{\rho_t}{\rho_P} \right)^{\frac{6\nu-2}{3\mu}} \right]^{\frac{2+\mu}{2}} \right]^{\frac{-3\mu}{2+\mu}} \tag{15}$$

with $K_1$=0.2, $K_2$=0.75, $K_r$=1.1, $\mu$=0.4, $\nu$=0.333 and Y = 1000 Pa. The value of the gravity on the lunar surface is g = 0.162 m s$^{-2}$; the parameters a, M, and V are the impactor radius, mass, and impact velocity, respectively. For meteoroid bulk densities $\rho_p$ of 0.3, 1.8 and 3.7 g cm$^{-3}$, Eq. (14) yields for the rim-to-rim crater diameter D 10.4 ± 0.5 m, 13.3 ± 0.6 m, and 15.8 ± 0.7 m, respectively, for a sporadic meteoroid hitting the Moon with an average collision velocity of 17 km s$^{-1}$.

Values derived from our analysis of the crater diameter are summarized in Table 3. Both above-mentioned scaling models predict a similar rim-to-rim diameter D for the same impactor bulk density, with D ranging from about 10 to 15 m. Because of its small size, this crater cannot be observed by telescopes from our planet. But probes in orbit around the Moon can spot it, provided that these can take pre- and post- impact images of the area where the meteoroid collision takes place. For instance, craters produced by previous collisions that gave rise to observed impact flashes were successfully identified by cameras onboard the Lunar Reconnaissance Orbiter (LRO), which orbits the Moon in a polar orbit since 2009 (Madiedo et al. 2014, 2019; Suggs et al. 2014, Robinson et al. 2015). These observations are or a paramount importance, since they would allow us to compare the actual and predicted crater diameters to check the validity of our assumptions. This would also provide a better constraint for the luminous efficiency associated with the collision of meteoroids on the Moon.

## 5. CONCLUSIONS

We have focused here on a lunar impact flash recorded during the Moon eclipse that occurred on 2019 January 21. This is the first impact flash unambiguously recorded on the Moon during a lunar eclipse and discussed in the scientific literature. The event, spotted and confirmed in the framework of the MIDAS survey, was also imaged by casual observers in Europe, America and Africa. The peak V magnitude of the flash was 4.2 ± 0.2, and its duration was of 0.28 s. According to our analysis, the most likely scenario with a probability of 99% is that the impactor that generated this flash was a sporadic meteoroid. By considering a value for the luminous efficiency of $3 \cdot 10^{-3}$ and an impact speed of 17 km/s, the estimated mass of the impactor yields 45 ± 8 kg. By employing the Gault scaling law, the rim-to-rim diameter of the crater generated during this collision ranges from 10.1 ± 0.5 m (for an impactor bulk density of 0.3 g cm$^{-3}$) to 15.3 ± 0.7 m (for a bulk density of 3.7 g cm$^{-3}$). The Holsapple model predicts a similar size. The crater could be measured by a probe in orbit around the Moon, such as for instance the Lunar Reconnaissance Orbiter. The comparison between the predicted and the experimental crater size could be very valuable to allow for a better constraint of the luminous efficiency for meteoroids impacting the lunar ground.

This is also the first time that lunar impact flash observations in more than two wavelengths are reported. The impact plume blackbody temperature has been estimated by analyzing the R, G and B channels of the color camera employed to record the event. This multiwavelength analysis has resulted in a peak temperature of 5700 ± 300 K.

ACKNOWLEDGEMENTS

We acknowledge funding from MINECO-FEDER project AYA2015-68646-P, and also from project J.A. 2012-FQM1776 (Proyectos de Excelencia Junta de Andalucía).

TABLES

| | |
|---|---:|
| Date and time | 2019 January 21 at 4h 41m 38.09 ± 0.01s UT |
| Peak brightness (magnitude) | 4.2 ± 0.2 in V band |
| Impact location | Lat.: 29.2 ± 0.3 °S, Lon.: 67.5 ± 0.4 °W |
| Duration (s) | 0.28 |
| Impactor kinetic energy (J) | $(6.55 \pm 0.63) \cdot 10^9$ |
| Impactor mass (kg) | 45 ± 8 |

Table 1. Characteristics of the lunar impact flash analysed here.

| | |
|:---:|:---:|
| $ZP_R + K_R A$ | 10.81 ± 0.06 |
| $ZP_V + K_V A$ | 11.07 ± 0.01 |
| $ZP_B + K_B A$ | 11.71 ± 0.02 |
| $C_R$ | -0.398 ± 0.11 |
| $C_V$ | -0.018 ± 0.006 |
| $C_B$ | 0.157 ± 0.05 |

Table 2. Results obtained from the photometric calibration of the Sony A7S camera, as defined by Equations (4 to 6).

| Scaling law | Impact angle (º) | Meteoroid Density (g cm$^{-3}$) | Meteoroid Mass (kg) | Impact Velocity (km s$^{-1}$) | Crater Diameter (m) |
|---|---|---|---|---|---|
| Gault | 45 | 0.3 | 45±8 | 17 | 10.1±0.5 |
| Gault | 45 | 1.8 | 45±8 | 17 | 13.6±0.6 |
| Gault | 45 | 3.7 | 45±8 | 17 | 15.3±0.7 |
| Holsapple | 45 | 0.3 | 45±8 | 17 | 10.4±0.5 |
| Holsapple | 45 | 1.8 | 45±8 | 17 | 13.3±0.6 |
| Holsapple | 45 | 3.7 | 45±8 | 17 | 15.8±0.7 |

Table 3. Diameter of the fresh crater, according to the Gault and the Holsapple models.

FIGURES

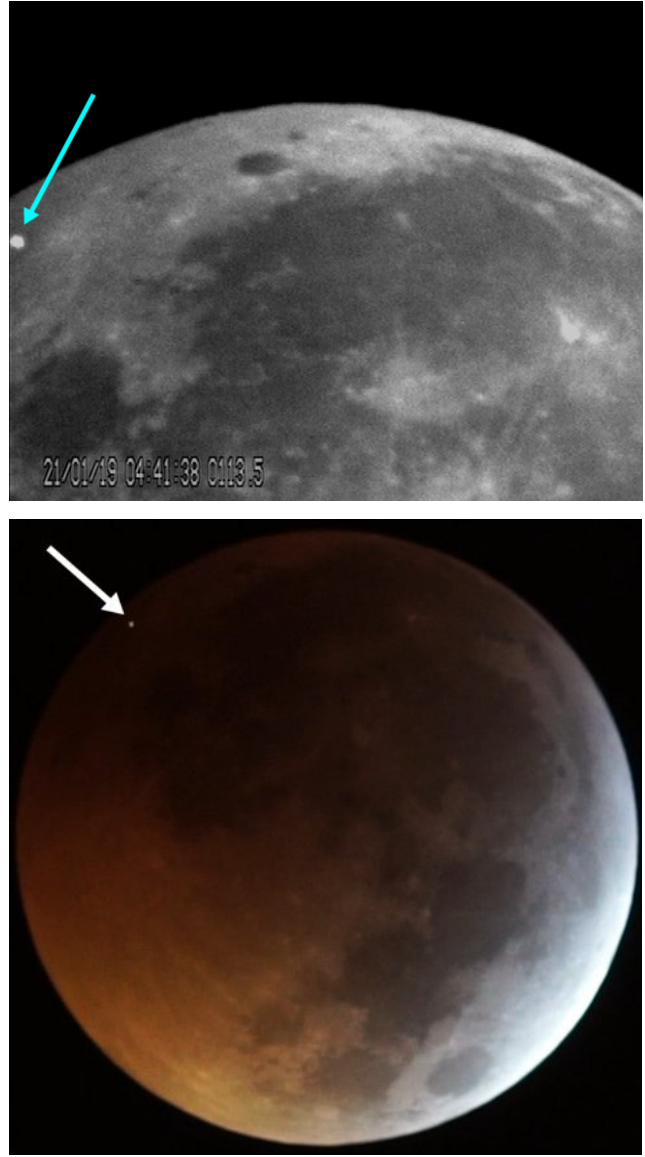

Figure 1. Lunar impact flash recorded on 2019 January 21 by the 0.36 m SC (up) and the 0.10 m refractor (down) telescopes.

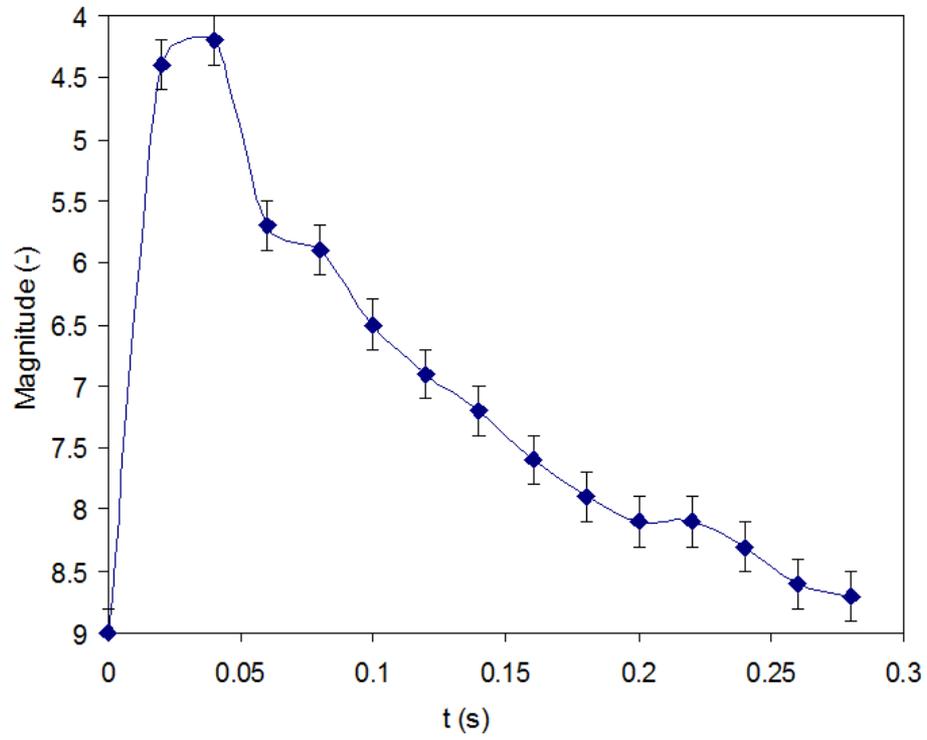

Figure 2. Lightcurve (evolution of V-magnitude as a function of time) of the impact flash recorded by the 0.36 m telescope.

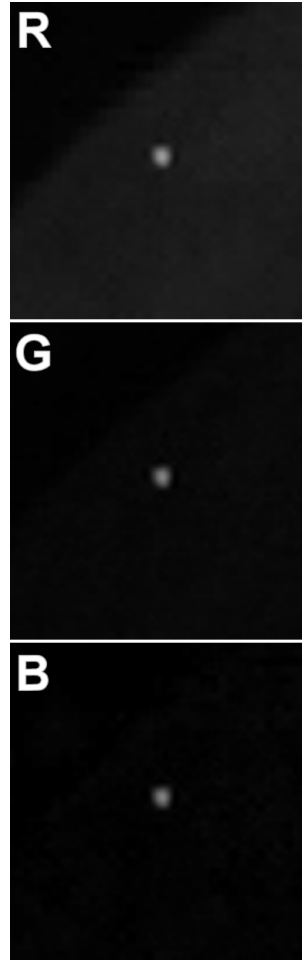

Figure 3. Decomposed image of the lunar impact flash into the three basic colour channels R, G, and B, during the peak luminosity of the event.

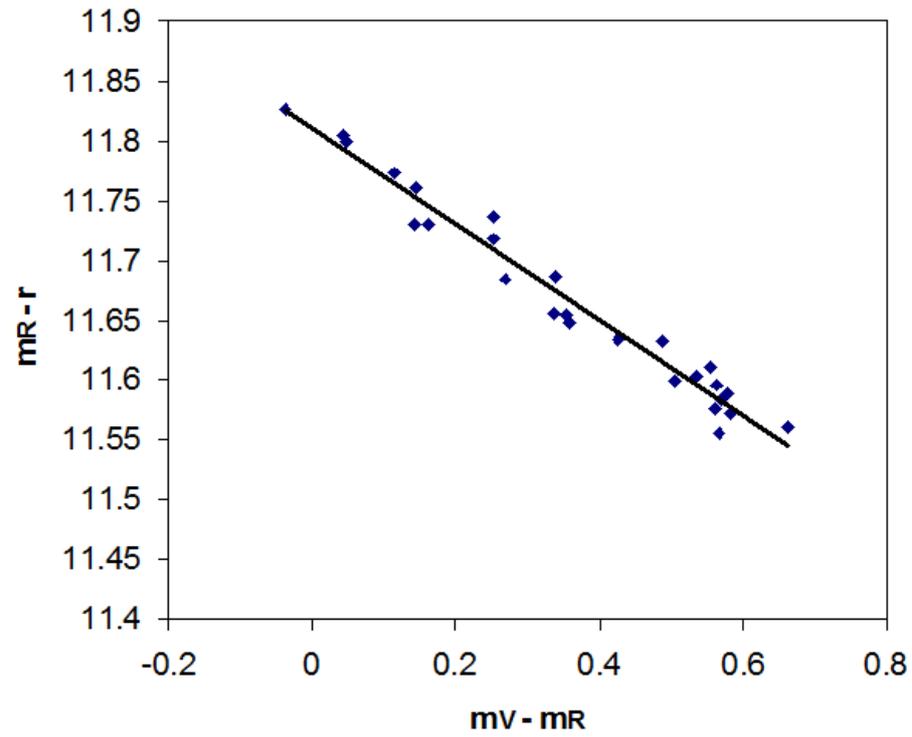

Figure 4. Photometric calibration for R band performed by employing 30 reference stars in Messier 67. The solid line corresponds to the best fit obtained from measured data.

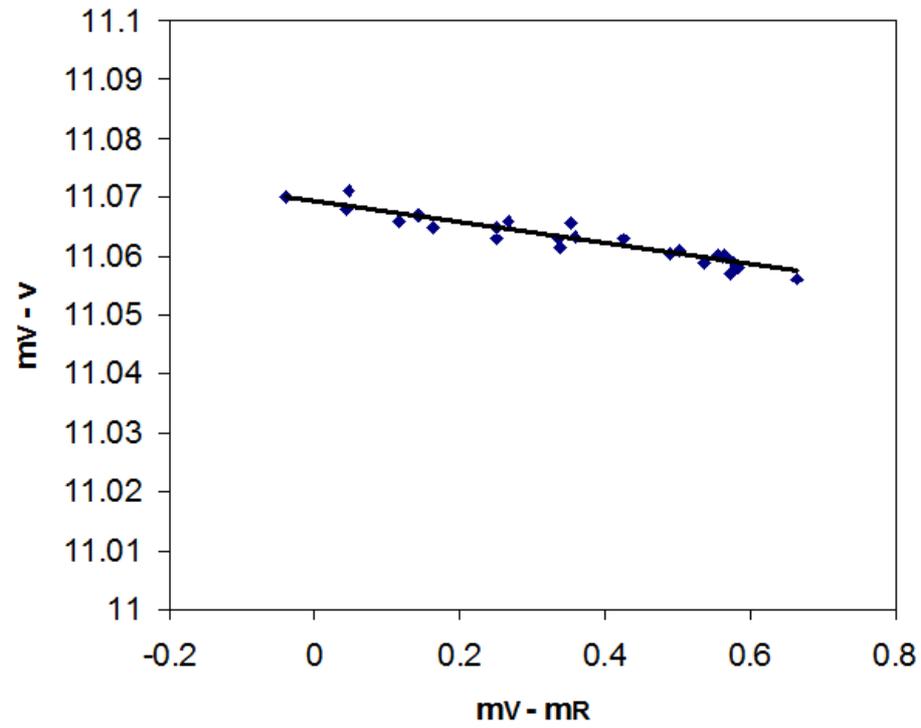

Figure 5. Photometric calibration for V band performed by employing 30 reference stars in Messier 67. The solid line corresponds to the best fit obtained from measured data.

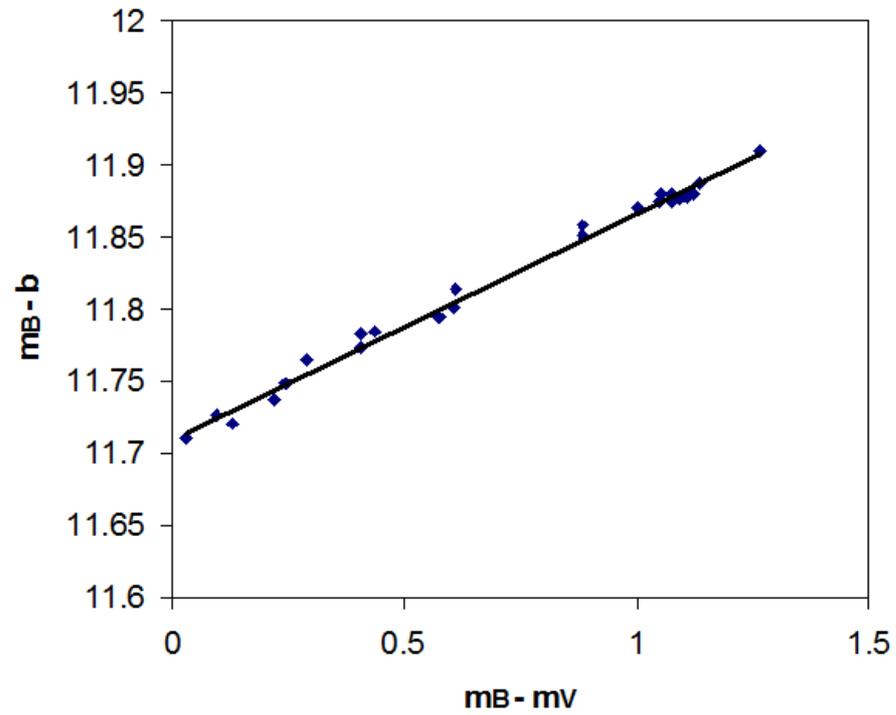

Figure 6. Photometric calibration for B band performed by employing 30 reference stars in Messier 67. The solid line corresponds to the best fit obtained from measured data.

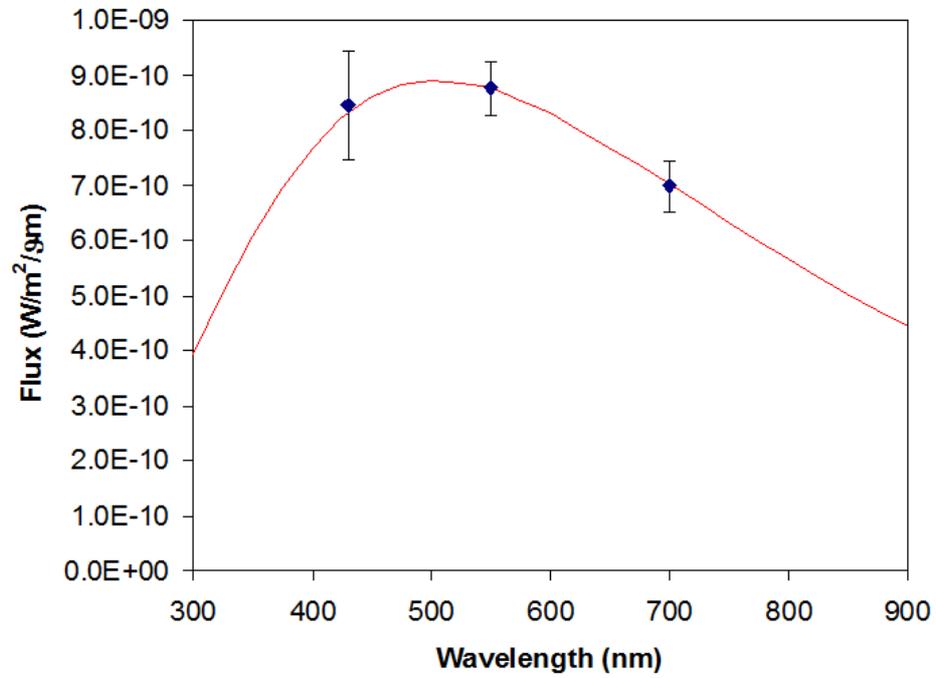

Figure 7. Flux densities obtained in R, V, and B bands. The solid line represents the best fit of these data to the flux emitted by a blackbody at a temperature T=5700 K.